\RequirePackage{fix-cm}
\documentclass[twocolumn,epjc3]{svjour3}  
\RequirePackage{graphicx,amssymb,amsfonts,amsmath,geometry,hyperref,color,lipsum}
\usepackage[english]{babel} 
\usepackage[T1]{fontenc}
\usepackage{pdflscape}
\usepackage{color}
\journalname{Eur. Phys. J. C}

\begin{document}
\title{{\bf {Bayesian Comparison of Interacting Modified Holographic Ricci Dark Energy Scenarios}}}
\author{Antonella Cid\thanksref{e1,addr1}
        \and
        Carlos Rodriguez-Benites\thanksref{e2,addr2}
        \and
        Mauricio Cataldo\thanksref{e3,addr1} \and
        Gonzalo Casanova\thanksref{e4,addr3}
}

\thankstext{e1}{e-mail: acidm@ubiobio.cl}
\thankstext{e2}{e-mail: carlos.rodriguez.b@uni.pe}
\thankstext{e3}{e-mail: mcataldo@ubiobio.cl}
\thankstext{e4}{e-mail: gonzalocasanova@udec.cl}

\institute{Departamento de F\'isica, Grupo Cosmolog\'ia y Part\'iculas Elementales, Universidad del B\'io-B\'io, Casilla 5-C, Concepci\'on, Chile.\label{addr1} \and
Universidad C\'esar Vallejo and Escuela Central de Posgrado, Universidad Nacional de Ingenier\'ia, Per\'u.
          \label{addr2}
           \and
Departamento de F\'isica, Universidad de Concepci\'on,  Casilla 160-C, Concepci\'on, Chile\label{addr3}
}

\date{Received: date / Accepted: date} 
\maketitle

\maketitle

\begin{abstract}
We perform a Bayesian model selection analysis for interacting scenarios of dark matter and modified holographic Ricci dark energy (MHRDE) with linear interacting terms. We use a combination of some of the latest cosmological data such as type Ia supernovae, cosmic chronometers, cosmic microwave background and baryon acoustic oscillations measurements. We find strong evidence against all the MHRDE interacting scenarios studied with respect to $\Lambda$CDM when the full joint analysis is considered.
\end{abstract}

\section{Introduction}
It is well known that our universe is currently in a phase of accelerated expansion  \cite{Weinberg:2012es}. This acceleration is driven by the so called dark energy, which in the standard cosmological model is represented by a cosmological constant $\Lambda$. The standard cosmological model or $\Lambda$CDM provides a good explanation for the current acceleration but it has some drawbacks, the cosmological constant problem \cite{Weinberg:1988cp,Weinberg:2000yb}, the coincidence problem \cite{Chimento:2003iea,Guo:2004vg,Pavon:2005yx} and the tension in the values obtained for the Hubble parameter from local measurements and inferred from Planck's data \cite{Verde:2019ivm,Riess:2016jrr,Riess:2011yx}.

Over the last twenty years, many dark energy models have been proposed in order to explain the observed current acceleration of the universe (see references \cite{Copeland:2006wr,Yoo:2012ug,Bahamonde:2017ize} for reviews on this topic). In particular, holographic dark energy models are based on the holographic principle (see reference \cite{Wang:2016och} for a review). According to \cite{Cohen:1998zx} the energy contained in a region of size $L$ must not exceed the mass of a black hole of the same size, i.e., in terms of the energy density $\rho\le L^{-2}$. In a cosmological context, the largest $L$ allowed is the one saturating this inequality. Based on this idea, Li \cite{Li:2004rb,Li:2010cj} proposed a model where the dark energy density is given by $\rho_x =3\tilde{c}^2 H^2$, where $\tilde{c}^2$ is a constant. 
Nevertheless, it is not possible obtain accelerated expansion from a model with the dark energy given by $\rho_x\propto H^2$ \cite{Hsu:2004ri}, because of this, alterative models motivated by the holographic principle have been explored. Among these, the model $\rho_x\propto R$ was proposed \cite{Gao:2007ep}, where $R$ is the Ricci scalar. Subsequently, in reference \cite{Cai:2008nk} it was noticed that the Ricci scalar curvature gives a causal connection scale of perturbations in the universe. There are many studies on these kind of models dubbed holographic Ricci dark energy (HRDE), e.g., see references \cite{Xu:2008rp,Feng:2008kz,Feng:2008rs,Li:2009bn,Zhang:2009un,Lepe:2010vh,Kim:2008ej,delCampo:2011jp,delCampo:2013hka}. Furthermore, in reference \cite{Granda:2008dk} an extension or modified holographic Ricci dark energy (MHRDE) model was proposed, where the dark energy density is given by 
\begin{equation}
\rho_x= 3(\alpha H^2+\beta \dot{H})\label{EOTH}  
\end{equation}
for $\alpha$ and $\beta$ constants. For more details on this model see references \cite{Granda:2008tm,Granda:2009xu,Granda:2009di,Wang:2010kwa,Mathew:2012md}.

On the other hand, in reference \cite{Hu:2006ar} the authors investigate a general formalism for interacting holographic dark energy models in order to solve the coincidence problem. In this work the characteristic size of holographic bound $L$ and the coupling term of interaction $Q$ for dark energy are not necessarily fixed. Over the years, many interacting holographic scenarios have been studied, see for example references \cite{Fu:2011ab,Chimento:2011dw,Chimento:2011pk,Chimento:2012zz,Chimento:2012hn,Chimento:2013se,P:2013cmq,Arevalo:2013tta,Chimento:2013qja,Oliveros:2014kla,Som:2014hja,Mahata:2015nga,Pan:2014afa,Arevalo:2014zoa,Lepe:2015qhq,Zadeh:2016vgc,Herrera:2016uci,Feng:2016djj,George:2019vko}.

In particular, scenarios with linear interaction, where the components are dark matter $\rho_m$ and holographic dark energy $\rho_x$, with interaction terms of the type $Q\propto H \rho_m$,  $Q\propto H \rho_x$ and $Q\propto H(\rho_m+\rho_x)$ are studied in~\cite{Fu:2011ab,P:2013cmq,Mahata:2015nga}. 
Interacting terms proportional to the dark energy densities and/or its derivatives in the context of modified holographic dark energy were studied in references \cite{Chimento:2011dw,Chimento:2012zz,Chimento:2012hn,Chimento:2013se}. Likewise, there are several models of non-linear interaction for dark matter and holographic dark energy, e.g., interacting terms  $Q\propto H\left(\frac{\rho^2_x}{\rho_m+\rho_x}\right)$, $Q\propto \left(\frac{\rho_m\rho_x}{H}\right)$ and $Q\propto H\left(\frac{\rho_m\rho_x}{\rho_m+\rho_x}\right)$ have been studied in references \cite{Oliveros:2014kla}, \cite{Mahata:2015nga,Pan:2014afa} and \cite{Feng:2016djj}, respectively.

In references \cite{Li:2009bn,delCampo:2011jp,Wang:2010kwa,Fu:2011ab,Feng:2016djj,George:2019vko,Akhlaghi:2018knk} the performance of holographic dark energy models in fitting the data has been compared with the $\Lambda$CDM model. In this sense, several criteria has been used, $\chi^2/dof$, AIC and BIC \cite{Arevalo:2016epc} and bayesian evidence. In this sense, bayesian model selection through the bayesian evidence is a more powerful statistical tool in comparing the performance of cosmological models in light of the more recent available data and it has been widely used in cosmology \cite{Santos:2016sti,Heavens:2017hkr,SantosdaCosta:2017ctv,Andrade:2017iam,Ferreira:2017yby}. In particular, in reference \cite{Cid:2018ugy} inconclusive evidence was found in studying a class of interacting models when compared to $\Lambda$CDM and considering background data.

The aim of this paper is to analyze the observational viability of interacting scenarios considering modified holographic Ricci dark energy. To asses the models' viability we perform a bayesian model selection analysis and compare interacting MHRDE scenarios with the $\Lambda$CDM model in light of background data such as supernovae type Ia, cosmic chronometers, baryon acoustic oscillations and the angular scale of the sound horizon at the last scattering. The paper is organized as follows. In section \ref{interaction} we find analytical solutions for the studied scenarios and describe the kinematics of these models. In section \ref{data} we describe the data used and the methodology. In section \ref{results} we discuss the main results and in section \ref{remarks} we present the final remarks.
%%%%%%%%%%%%%%%%%%%%%%%%%%%%%%%%%%%%%%%%%%%%%%%%%%%%%%%%%%%%%%%%%%%%%%%%%%%%%%%%%%%%%%%%%%%%%%%%%%%%%%%%%%%%%%%%%%%%%%%%%%%%%%%%%%%%%%%%%%%%%
\section{The interacting modified holographic dark energy model}\label{interaction}
Let us consider a flat, homogeneous and isotropic universe in the framework of General Relativity, the spatially flat Friedmann-Lema\^itre-Robertson-Walker (FLRW) metric. 
The Friedmann's equation in this context is written as
\begin{eqnarray}\label{EEH1}
3\, H^2 &=& \rho\, ,
\end{eqnarray}
where $\rho$ is the total energy density and $8\pi G=c=1$ is assumed. On the other hand, from the conservation of the total energy-momentum tensor we have 
\begin{equation}\label{EcuaciondeConservacion}
\dot{\rho}+3H(\rho+p)=0\,,
\end{equation}
where $p$ is the total pressure. 
A realistic cosmological scenario contains baryons ($b$), radiation ($r$), cold dark matter ($c$) and a dark energy ($x$) components, in this work this last component is assumed to be given by holographic dark energy. In this context we consider the Friedmann equation~\eqref{EEH1} and the conservation equation~\eqref{EcuaciondeConservacion} assuming $\rho= \rho_b+\rho_r+\rho_c+\rho_x$ and $p= p_b+p_r+p_c+p_x$. From here on and for the sake of simplicity we define $\rho_d=\rho_c+\rho_x$. In addition, a barotropic equation of state is considered for all the components, $p_i=\omega_i\,\rho_i$ with $\omega_b=0$, $\omega_r=1/3$, $\omega_c=0$ and $\omega_x=\omega$ as a state function. Furthermore, by including a phenomenological interaction in the dark sector $\Gamma$, we separate the conservation equation~\eqref{EcuaciondeConservacion} into the following equations
\begin{eqnarray}
\rho'_b+\rho_b&=&0,\label{4FluidosEC1'}\\
\rho'_r+\frac{4}{3}\rho_r &=&0,\label{4FluidosEC2'}\\
\rho'_c+\rho_c &=&-\Gamma,\label{4FluidosEC3'}\\
\rho'_x+(1+\omega)\,\rho_x &=&\Gamma,\label{4FluidosEC4'}
\end{eqnarray}
where by convenience we use the change of variable $\eta=3\ln a$ and define $(\,)'= d/d\eta$.
Note that $\Gamma > 0$ indicates an energy transfer from cold dark matter to dark energy and $\Gamma < 0$ indicates the opposite. From Eqs.~\eqref{EOTH} and~\eqref{EEH1} we can easily notice that  
\begin{equation}\label{4FluidosEOTH}
\rho_x=\alpha\,\rho+\frac{3\beta}{2}\,\rho'\,.
\end{equation}
By deriving Eq.~\eqref{4FluidosEOTH} and replacing $\rho_x'=\rho_d'-\rho_c'$, $\rho_c'$ from Eq. \eqref{4FluidosEC3'}, $\rho_c=\rho_d-\rho_x$, $\rho_x$ from Eq.~\eqref{4FluidosEOTH},
$\rho=\rho_{b}+\rho_{r}+\rho_d$, $\rho''_b=\rho_b$, $\rho''_r=\frac{16}{9}\rho_r$ and $\rho_r$ by the solution of Eq.~\eqref{4FluidosEC2'}, in this order, we obtain a second order differential equation for the energy density of the dark sector $\rho_d$, 
\begin{eqnarray}
\frac{3\beta}{2}\,\rho''_d+\left( \alpha+\frac{3\beta}{2}-1 \right)\,\rho'_d+(\alpha-1)\,\rho_d\nonumber \\
+\frac{1}{3}(2\beta-\alpha)\,\rho_{r0}\,a^{-4}=\Gamma,\label{4FluidosEDRho}
\end{eqnarray}	
where $\rho_{r0}$ is the integration constant from Eq.~\eqref{4FluidosEC2'}. For a given interaction $\Gamma=\Gamma(\rho_d,\rho'_d,\rho,\rho')$, we can analytically solve Eq.~\eqref{4FluidosEDRho} to find the energy density $\rho_d$, and consequently the energy densities $\rho_x$ and $\rho_c$ through Eq.~\eqref{4FluidosEOTH}. 

In this work we study the general linear interaction,
\begin{eqnarray}
\label{Int_def}
\Gamma=\alpha_1\rho_c+\beta_1\rho_x,
\end{eqnarray}
which includes four different types of interaction, $\alpha_1=0$, $\beta_1=0$, $\alpha_1=0$, $\alpha_1=\beta_1$ and $\alpha_1\neq\beta_1$. Notice that it is possible to describe all these interactions with terms proportional to $\rho_d$, $\rho'_d$, $\rho$ and $\rho'$. Then, we can  rewrite Eq.~\eqref{4FluidosEDRho} as
\begin{equation}
\rho''_d+b_1\,\rho'_d+b_2\,\rho_d+b_3\,a^{-3}+b_4\,a^{-4}=0\,,\label{4FluidosEDRhoGeneral}
\end{equation}
including the four interaction types of our interest, where the constants are defined as
\begin{eqnarray*}
b_1&=&1+\alpha_1-\beta_1-2(1-\alpha)/3\beta,\\ b_2&=&\frac{2}{3\beta}\left(\alpha(1-\beta_1+\alpha_1)-1-\alpha_1\right),\\ 
b_3&=&\Omega_{b0}(\beta_1-\alpha_1)\left( 1-2\alpha/{3\beta}\right) ,\\ 
b_4&=&\frac{2\Omega_{r0}}{3\beta} \left((2\beta-\alpha)/3-(\beta_1-\alpha_1)(\alpha-2\beta)\right),
\end{eqnarray*}
and $\Omega_{b0}$ and $\Omega_{r0}$ are the density parameters (i.e. $\Omega_{i0}=\rho_{i0}/3H^2_0$ with $i=\{b,r\}$ for baryons, the radiation and $H_0$ is the Hubble parameter). The general solution of equation~\eqref{4FluidosEDRhoGeneral} has the form
\begin{equation}\label{4FluidosEDRhoSolGeneral}
\rho_d(a)=3H_0^2\left(\frac{A}{a^3} +\frac{B}{a^4}+C_1\,a^{3\,\lambda_1}+C_2\,a^{3\,\lambda_2}\right)\,,
\end{equation}
where the integration constants are given by
\begin{eqnarray}
&&C_1= \frac{A(1+\lambda_2)}{(\lambda_1-\lambda_2)}+\frac{B(4+3\lambda_2)}{3(\lambda_1-\lambda_2)}+\frac{2\,(\Omega_{x0}-\alpha)}{3\beta(\lambda_1-\lambda_2)}\nonumber \\
&&+\frac{3\Omega_{b0}+4\Omega_{r0}-3\lambda_2(\Omega_{c0}+\Omega_{x0})}{3(\lambda_1-\lambda_2)}\,,\\
&&C_2=-A-B+\Omega_{c0}+\Omega_{x0}-C_1\,,
\end{eqnarray}
and $\Omega_{c0}$, $\Omega_{x0}$ are the density parameters for the cold dark matter and the MHRDE, respectively. The coefficients in~\eqref{4FluidosEDRhoSolGeneral} are given by,
\begin{eqnarray}
&&A = \frac{b_3}{b_1-b_2-1}\,, \,\,B = \frac{9b_4}{12b_1-9b_2-16}\,,\nonumber\\
&&\textrm{and}\quad \lambda_{1,2}=-\frac{1}{2}\left(b_1\pm\sqrt{b^2_1-4b_2}\right)\,.\label{4FluidosValoresdeLambdas}
\end{eqnarray}
Therefore, the Hubble expansion rate can be written as:
\begin{equation}\label{Hz}
H(a)^2=H_0^2\left(\frac{\bar{A}}{a^3}+\frac{\bar{B}}{a^4}+C_1a^{3\lambda_1}+C_2a^{3\lambda_2}\right),
\end{equation}
where $\bar{A}=A+\Omega_{b0}$, $\bar{B}=B+\Omega_{r0}$, $\Omega_{b0}+\Omega_{r0}+\Omega_{c0}+\Omega_{x0}=1$ and the radiation term includes the contribution of photons, $\Omega_{\gamma 0}$, and neutrinos, $\Omega_{\nu 0}$.

Notice that, without an interacting term, a HRDE model, $\rho_x=\alpha(2H^2+\dot{H}$), is recovered from \eqref{Hz} for $\alpha=2\beta$, $b_3=b_4=0$ and $A=B=0$. Likewise, a MHRDE model, $\rho_x=3(\alpha H^2+\beta\dot{H}$), is recovered from \eqref{Hz} for $b_3=0$ and $A=0$.

%%%%%%%%%%%%%%%%%%%%%%%%%%%%%%%%%%%%%%%%%%%%%%%%%%%%%%%%%%%%%%%%%%%%%%%%%%%%%%%%%%%%%%%%%%%%%%%%%%%%%%%%%%%%%%%%%%%%%%%
On the other hand, using Eqs.~\eqref{4FluidosEOTH},~\eqref{Int_def} and ~\eqref{4FluidosEDRhoSolGeneral} into~\eqref{4FluidosEC4'}, we obtain an expression for the variable state parameter,
\begin{equation}\label{4FluidosWcomofunciondea}
\omega(a)=\frac{D_1\, a^{-3} +D_2\, a^{-4}+D_3\,a^{3\,\lambda_1}+D_4\,a^{3\lambda_2}}{\tilde{A}\, a^{-3} + \tilde{B}
	\, a^{-4}+\tilde{C}_1\,a^{3\,\lambda_1}+\tilde{C}_2\,a^{3\lambda_2}}\, , 
\end{equation}
where $\tilde{A}=(2\alpha-3\beta)\bar{A}$, $\tilde{B}=2(\alpha-2\beta)\bar{B}$, $ \tilde{C}_1=C_1(3\beta\lambda_1+2\alpha)$, $\tilde{C}_2=C_2(3\beta\lambda_2+2\alpha)$, $D_1=2\alpha_1A+(\beta_1-\alpha_1)\tilde{A}$, $D_2=2\alpha_1B+(1/3-\alpha_1+\beta_1)\tilde{B}$, $D_3=-2C_1(1+\lambda_1)$ and $D_4=-2C_2(1+\lambda_2)$. In the limit to the future ($a\rightarrow \infty$), the expression~\eqref{4FluidosWcomofunciondea} becomes $\omega\rightarrow\frac{-2(1+\lambda_2)}{3\beta\lambda_2+2\alpha}$ for $\lambda_2> \lambda_1>-1$, which could assume positive or negative values depending on the interacting and holographic parameters.

In addition, from~$\rho_c=\rho_d-\rho_x$,~\eqref{4FluidosEOTH} and~\eqref{4FluidosEDRhoSolGeneral}, the coincidence parameter ($r=\rho_c/\rho_x$) becomes,
\begin{eqnarray}\label{Solder}
r(a)=\frac{\hat{A}a^{-3}+\hat{B}a^{-4}+\hat{C}_1a^{3\lambda_1}+\hat{C}_2a^{3\lambda_2}}{\tilde{A}\, a^{-3} + \tilde{B}
	\, a^{-4}+\tilde{C}_1\,a^{3\,\lambda_1}+\tilde{C}_2\,a^{3\lambda_2}}
\end{eqnarray}
where $\hat{X}=2X-\tilde{X}$ for $X=A,B,C_1,C_2$. Therefore, the asymptotic limit of $r(a)$ as $a \rightarrow \infty$ for $\lambda_2> \lambda_1>-1$ becomes
\begin{equation}\label{4FluidosSolrinfinito}
r_{\infty}\rightarrow\frac{2}{3\beta\lambda_2+2\alpha}-1\,,
\end{equation}
a constant depending on the interacting and holographic parameters.

Furthermore, from Eq.~\eqref{4FluidosEDRhoSolGeneral} we can obtain the deceleration parameter $q=-\left(1+\frac{3\rho'}{2\rho}\right)$ as
\begin{equation}
q(a)=-1+\frac{3}{2}\frac{\frac{\bar{A}}{a^{3}}+\frac{4}{3}\frac{\bar{B}}{a^{4}}-C_1\lambda_1a^{3\lambda_1}-C_2\lambda_2a^{3\lambda_2}}{\frac{\bar{A}}{a^{3}}+\frac{\bar{B}}{a^{4}}+C_1a^{3\lambda_1}+C_2a^{3\lambda_2}}\,.
\end{equation}
Notice that in the asymptotic limit $a\rightarrow \infty$ we get $q\rightarrow -1-\frac{3}{2}\lambda_2$ for  $\lambda_2> \lambda_1>-1$.

%%%%%%%%%%%%%%%%%%%%%%%%%%%%%%%%%%%%%%%%%%%%%%%%%%%%%%%%%%%%%%%%%%%%%%%%%%%%%%%%%%%%%%%%%%%%%%%%%%%%%%%%%%%%%%%
\section{Observational analysis}\label{data}

In the observational analysis we use data such as cosmic chronometers, obtained through the differential age method and reported in reference~\cite{Moresco:2016mzx}, supernovae type Ia (SNe Ia) from the Pantheon Sample~\cite{Scolnic:2017caz}, baryon acoustic oscillations from 6dFGS~\cite{Beutler:2011hx}, SDSS-MGS~\cite{Ross:2014qpa}, eBOSS~\cite{Ata:2017dya},\cite{Hou:2018yny}, BOSS DR12~\cite{Alam:2016hwk} and BOSS Ly$\alpha$ \cite{Bourboux:2017cbm}, and the angular scale of the sound horizon at the last scattering~\cite{Ade:2015rim}. In the following, we briefly present each one of these dataset.
%%%%%%%%%%%%%%%%%%%%%%%%%%%%%%%%%%%%%%%%%%%%%%%%%%%%%%%%%%%%%%%%%%%%%%%%%%%%%%%%%%%%%%%%%%%%%%%%%%%%%%%%%%%%%%%%%
\subsection{Cosmic chronometers} 
We use 24 cosmic chronometers obtained through the differential age method (see table 4 in \cite{Moresco:2016mzx}) by taking the relative age of passively evolving galaxies with respect to the redshift~\cite{Jimenez:2001gg}. This procedure provides cosmological-independent direct measurements of the expansion history of the universe up to redshift 1.2~\cite{Verde:2014qea}.
In our analysis, the theoretical value of the Hubble expansion rate is given by equation~\eqref{Hz}. 

%%%%%%%%%%%%%%%%%%%%%%%%%%%%%%%%%%%%%%%%%%%%%%%%%%%%%%%%%%%%%%%%%%%%%%%%%%%%%%%%%%%%%%%%%%%%%%%%%%%%%%%%%%%%%%%%%
\subsection{Supernovae Type Ia}
We use the most up to date compilation of supernovae type Ia, the Pantheon Sample, containing a set of 1048 spectroscopically confirmed SNe Ia~\cite{Scolnic:2017caz} ranging from redshift $0.01$ to $2.3$, along with a covariance matrix (including statistical and systematic errors). The Pantheon catalog contains measurements of peak magnitudes in the B-band's rest frame, $m_B$, which are related to the distance modulus as $\mu=m_B+M_B$, where $M_B$ is a nuisance parameter corresponding to the absolute B-band magnitude of a fiducial SN Ia.
In our analysis we theoretically compute the distance modulus at a given redshift as
\begin{equation}\label{modulo_de_distancia}
\mu(z)=5\log d_L(z)+25,
\end{equation}
where $d_L$ is the luminosity distance in units of Mpc,
\begin{equation}
    d_L=(1+z)\int^z_0\frac{H_0\, dz'}{H(z')},
\end{equation}
and $H_0$ is the Hubble parameter.

%%%%%%%%%%%%%%%%%%%%%%%%%%%%%%%%%%%%%%%%%%%%%%%%%%%%%%%%%%%%%%%%%%%%%%%%%%%%%%%%%%%%%%%%%%%%%%%%%%%%%%%%%%%%%%%%%
\subsection{BAO data}
The isotropic measurements of the BAO signal are given in terms of the dimensionless ratio
\begin{equation}
    d_z(z) = D_V (z)/r_s(z_d)
\end{equation}
where $D_V$ is a combination of the line-of-sight and transverse distance scales defined in reference~\cite{Eisenstein:2005su}, $z_d$ is the redshift at the drag epoch and $r_s(z)$ is the comoving size of the sound horizon, where $D_V$ and $r_s$ are defined by
\begin{eqnarray}
    D_V(z)&=&\left((1 + z)^2D_A(z)^2 \frac{cz}{H(z)}\right)^{1/3} \, \text{and}\\ r_s(z)&=&\int^{\infty}_z \frac{c_s dz}{H(z)} 
\end{eqnarray}
respectively, with $c$ the speed of light, $D_A(z) = \frac{c}{(1+z)}\int^z_0 \frac{dz}{H(z)}$ the angular diameter distance, $c_s = \frac{c}{\sqrt{3(1+R)}} $ being the sound speed of the photon-baryon fluid and $R = \frac{3\Omega_b}{4\Omega_{\gamma}(1+z)}$~\cite{Eisenstein:1997ik}. 

We use isotropic BAO measurements from 6dFGS~\cite{Beutler:2011hx}, MGS~\cite{Ross:2014qpa} and eBOSS~\cite{Ata:2017dya}.

Furthermore, we use the anisotropic BAO measurements from BOSS DR12~\cite{Alam:2016hwk} and Ly$\alpha$ forest~\cite{Bourboux:2017cbm}, which are defined in terms of $D_A$ and $D_H=c/H(z)$, as shown in table 2 of Ref.~\cite{Evslin:2017qdn}. We use these data along with the corresponding covariance matrix in Ref.~\cite{Evslin:2017qdn}.
%%%%%%%%%%%%%%%%%%%%%%%%%%%%%%%%%%%%%%%%%%%%%%%%%%%%%%%%%%%%%%%%%%%%%%%%%%%%%%%%%%%%%%%%%%%%%%%%%%%%%%%%%%%%%%%%%
\subsection{CMB data}
We use the CMB compressed likelihood and fix the physical baryon density to $\Omega_bh^2 = 0.022383$, as reported in~\cite{Aghanim:2018eyx}.
The only contribution of CMB data we consider in this work is the angular scale of the sound horizon at the last scattering:
\begin{equation}
    \ell_a =\frac{\pi (1+z_*)D_A(z_*)}{r_s(z_*)}
\end{equation}
where the comoving size of the sound horizon is evaluated at $z_* = 1089.80$, according with Planck’s 2018 results~\cite{Aghanim:2018eyx}. We compare the value obtained in our study with the one reported by the Planck collaboration in 2015, $\ell_a=301.63\pm0.15$~\cite{Ade:2015rim}.

%%%%%%%%%%%%%%%%%%%%%%%%%%%%%%%%%%%%%%%%%%%%%%%%%%%%%%%%%%%%%%%%%%%%%%%%%%%%%%%%%%%%%%%%%%%%%%%%%%%%%%%%
\subsection{Bayesian model selection}

The Bayesian inference (based on the Bayes' theorem) is a robust statistical technique for parameter estimation and model selection widely used in the study of cosmological scenarios~\cite{Cid:2018ugy,Arevalo:2016epc}. The Bayes' theorem relates the posterior probability $P$ for a set of parameters $\Theta$, given the data $\mathcal{D}$, described by a model $\mathcal{M}$,
\begin{equation}\label{Probability}
    P(\Theta\mid \mathcal{D,M})=\frac{\mathcal{L}(\mathcal{D}\mid \Theta,\mathcal{M})\,\mathcal{P}(\Theta\mid \mathcal{M})}{\mathcal{E}(\mathcal{D}\mid \mathcal{M})}
\end{equation}
where $\mathcal{L}$, $\mathcal{P}$ and $\mathcal{E}$ are the likelihood, prior and evidence, respectively.

In comparing the performance of two different models given a dataset, we use the Bayes' factor defined as the ratio of the evidences of models $M_1$ and $M_2$ as:
\begin{equation}\label{Bij}
    B_{12}=\mathcal{E}_1/\mathcal{E}_2
\end{equation}
where the evidence is obtained integrating Eq. \eqref{Probability} over the space of parameters. If the models $M_1$ and $M_2$ have the same prior probability, then the Bayes factor gives the posterior odds of the two models.

To compare the studied models with the $\Lambda$CDM model, we use a conservative version of the Jeffreys' scale defined in reference~\cite{Trotta:2008qt}. This scale gives us an empirical measure for interpreting the strength of the evidence in comparing two competing models. The Jeffreys' scale interprets the evidence as follows, inconclusive if $\vert\ln B_{12}\vert<1$, weak if $1\le\vert\ln B_{12}\vert<2.5$, moderate if $2.5\le\vert\ln B_{12}\vert<5$ and strong if $\vert\ln B_{12}\vert\ge5$. In all the cases, the evidence is interpreted as in favor of the tested model if $\ln B_{12}$ is positive or against if negative.

In our work we consider $\Lambda$CDM as the reference model, as such, the subscripts in the Bayes’ factor~\eqref{Bij} will be omitted hereafter.

To compute the evidence and generate the posterior distributions we use the \textsc{MultiNest} algorithm\footnote{\url{https://github.com/JohannesBuchner/MultiNest}} \cite{Feroz:2007kg,Feroz:2008xx}, requiring a global log-evidence tolerance of 0.01 as a convergence criterion and working with a set of 800 live points to improve the accuracy in the estimation of the evidence. 

\section{Analysis and Results}\label{results}
We performed a Bayesian comparison analysis of the general interaction model $\Gamma=\alpha_1\rho_c+\beta_1\rho_x$ with the $\Lambda$CDM model in terms of the strength of the evidence according to the Jeffreys' scale. We consider a combination of background data, type Ia supernovae, cosmic chronometers, baryonic acoustic oscillations and cosmic microwave background.
 
In the studied models the prior probability distributions for free parameters  are shown in table~\ref{tab1}. We have chosen a uniform prior for parameters such as $\Omega_c$, $\alpha$, $\beta$, $\alpha_1$, $\beta_1$, and $M_B$, and a Gaussian prior for the parameter $h$. For the parameter $\Omega_c$ we choose a conservative uniform prior between 0 and 1, for the dimensionless Hubble parameter $h$ we adopt a Gaussian prior centered in the value reported by Riess et al. in Ref.~\cite{Riess:2018byc}. The priors for the holographic parameters were considered positive and small~\cite{Chimento:2012hn,Arevalo:2013tta}, the prior for the interacting parameters are uniform distributions centered in zero and for the Pantheon Sample parameter $M_B$, we use a conservative range including the value reported by Scolnic et al. in reference~\cite{Scolnic:2017caz}.

\begin{center}
\begin{table}[ht]
\caption{Prior probability distributions for the models' parameters. For a Gaussian (G) prior we inform $(\mu,\sigma^2)$ and for a Uniform (U) prior we inform $(a,b)$ representing $a\leq x \leq b$.}\label{tab1}
\centering
\begin{tabular}{|c|c|c|} 
\hline
 Parameters & Prior & Ref.\\
\hline
$h$ & G$(0.7352,0.0162)$&~\cite{Riess:2018byc}\\
$\Omega_c$ & U$(0,1)$ &-\\
$\alpha$ & U$(0,1)$ &~\cite{Chimento:2012hn,Arevalo:2013tta}\\
$\beta$ &  U$(0,1)$ &~\cite{Chimento:2012hn,Arevalo:2013tta}\\
$\alpha_1$& U$(-1,1)$ &~\cite{Arevalo:2016epc,Cid:2018ugy}\\
$\beta_1$ & U$(-1,1)$ &~\cite{Arevalo:2016epc,Cid:2018ugy}\\
$M_B$  & U$(-20,-18)$ &~\cite{Scolnic:2017caz}\\
\hline
\end{tabular}
\end{table}
\end{center}
 
We expect interacting models mainly affect the late time evolution and not the physics of the primordial universe. In this sense we consider the following constraints: $\Omega_b h^2=0.022383$ \cite{Aghanim:2018eyx}, $\Omega_r=\Omega_\gamma\left( 1+\frac{7}{8}(\frac{4}{11})^{\frac{4}{3}} N_{eff}\right)$ with $N_{eff}=3.046$~\cite{Mangano:2005cc}, and $\Omega_\gamma h^2=2.469\times 10^{-5}$~\cite{Komatsu:2010fb}. Moreover, for the redshift at the drag epoch and the last scattering epoch we use Planck's results~\cite{Aghanim:2018eyx}, $z_d = 1060.01$ and $z_* = 1089.80$, respectively.

Our main results are summarized in tables~\ref{tab2}-\ref{tab3}. We labeled the studied interacting modified holographic Ricci dark energy models (IMHRDE) in the following way, IMHRDE 1, 2, 3, 4 representing $\alpha_1=0$, $\beta_1=0$, $\alpha_1=\beta_1$, $\alpha_1\neq\beta_1$, respectively.

In  tables~\ref{tab2} and~\ref{tab3} we present the mean value and $1\sigma$ error for the parameters of all the studied models, along with the logarithm of the evidence, the logarithm of the Bayes factor and the interpretation for the strength of the evidence. We notice that by considering the Pantheon sample only we get weak evidence against for each of the studied IMHRDE when compared to $\Lambda$CDM (see table~\ref{tab2}).  In table~\ref{tab3} the results for the full joint analysis, including SNe Ia from Pantheon Sample, BAO, CC, and CMB are shown, here we find strong evidence against for each of the studied IMHRDE model.

As a comparison, we also indicate the results for models HRDE and MHRDE (without interaction). For these scenarios in the full joint analysis, evidence indicates more support when compare to the interacting ones, nevertheless the evidence also disfavor these models when compare to $\Lambda$CDM.

On the other hand, in table \ref{tab3} we notice that the best fit parameters for models IMHRDE 2 and IMHRDE 3 are consistent with those of MHRDE, this is because the interacting parameters are consistent with zero (at 1$\sigma$ level) in these cases, we further notice that the addition of interaction for these scenarios shift the evidence for these models from moderate for the model MHRDE to strong for the IMHRDE models. Thus, we conclude that the studied IMHRDE models are disfavored when compared to $\Lambda$CDM in a full joint analysis with background data and considering the Pantheon sample only. The evidence shift to a better support for $\Lambda$CDM when high-redshift data (CMB and BAO) are considered.

In the literature there are several studies analyzing the performance of holographic dark energy models in fitting background data compared to $\Lambda$CDM. In particular, in reference \cite{Li:2009bn} the holographic Ricci dark energy model (HRDE) without interaction is analyzed, the authors find evidence disfavoring this model when compared to $\Lambda$CDM. In reference \cite{delCampo:2011jp} interacting HRDE is studied with the AIC and BIC criteria and the interacting HRDE model is concluded ruled out. In reference \cite{Wang:2010kwa} a modified holographic Ricci dark energy model (MHRDE) without interaction is considered and the $\chi^2/dof$ criteria is used to reach the same conclusion. In reference \cite{Feng:2016djj} many interacting HRDE models are studied and all of them are discarded according to the BIC criteria when compared to $\Lambda$CDM. In reference \cite{Akhlaghi:2018knk} HRDE and MHRDE models are analyzed without interaction, beyond background data the authors consider growth rate data, with AIC and BIC the authors find strong indications against holographic models when compared to $\Lambda$CDM. Likewise, in our work we find strong evidence against the linear interacting modified holographic Ricci dark energy models studied (see table \ref{tab3}) when compared to $\Lambda$CDM in light of the most recent background data.

\begin{figure}[ht]
 \centering
 \includegraphics[width=0.50\textwidth]{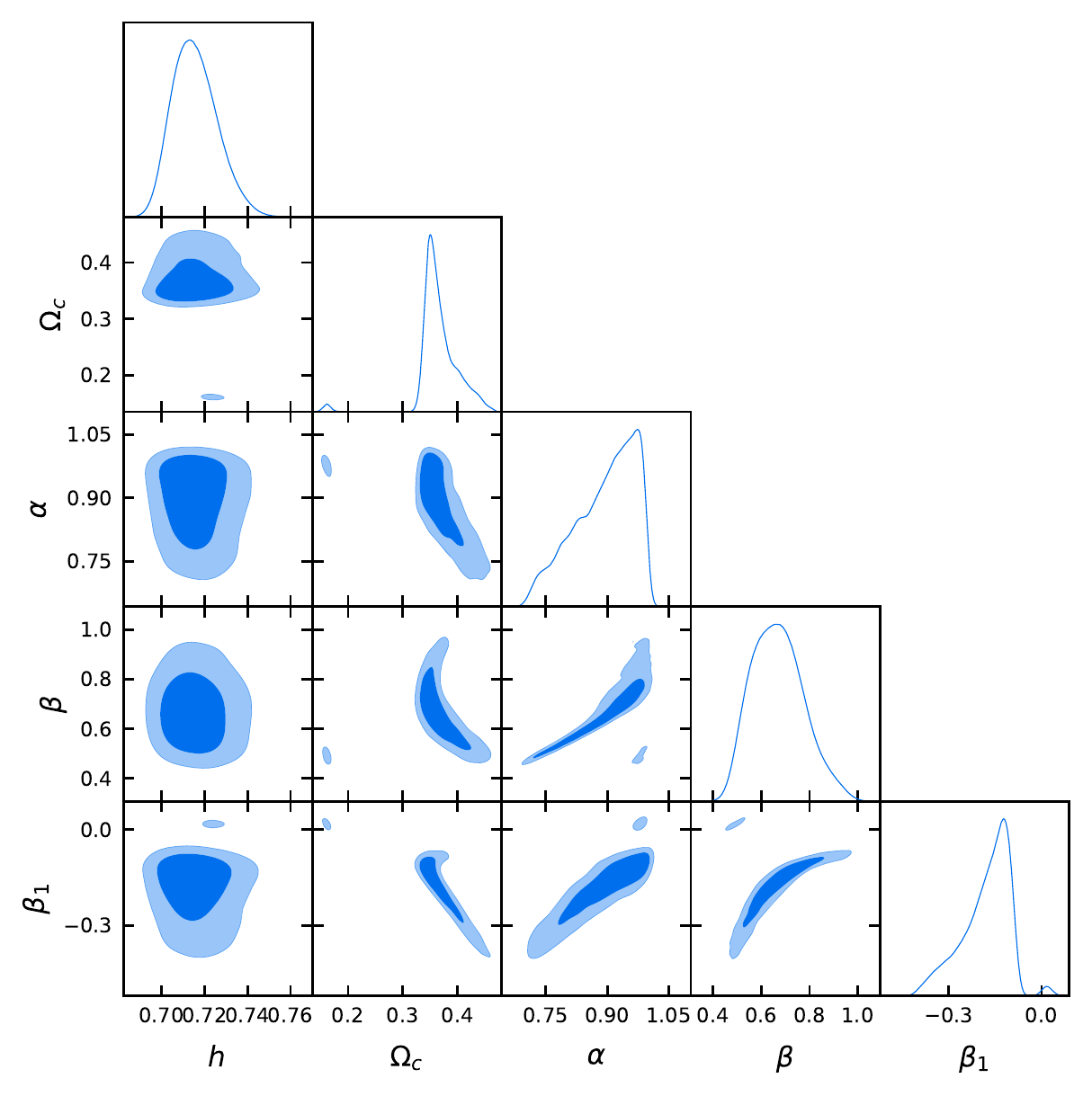}
 \caption{Contour plots for interacting model 1 showing the $1\sigma$ and $2\sigma$ regions. We considered the full joint analysis with Pantheon + BAO + CC + CMB.} \label{Model1}
 \end{figure}

\begin{figure}[ht]
\centering
\includegraphics[width=0.50\textwidth]{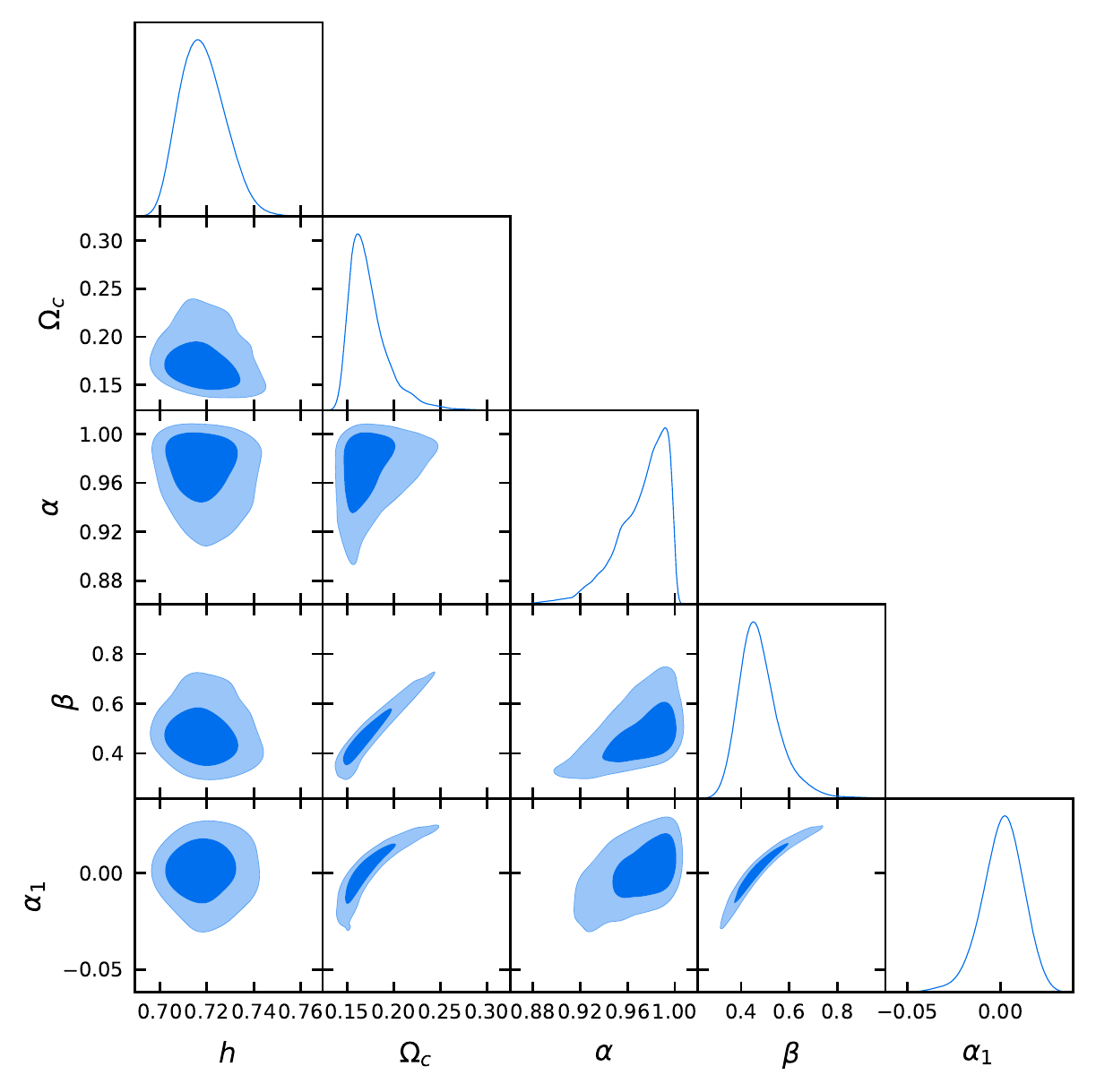}
\caption{Contour plots for interaction model 2 with the $1\sigma$ and $2\sigma$ regions. We considered the full joint analysis with Pantheon + BAO + CC + CMB.} \label{Model2}
\end{figure}

\begin{figure}[ht]
\centering
\includegraphics[width=0.50\textwidth]{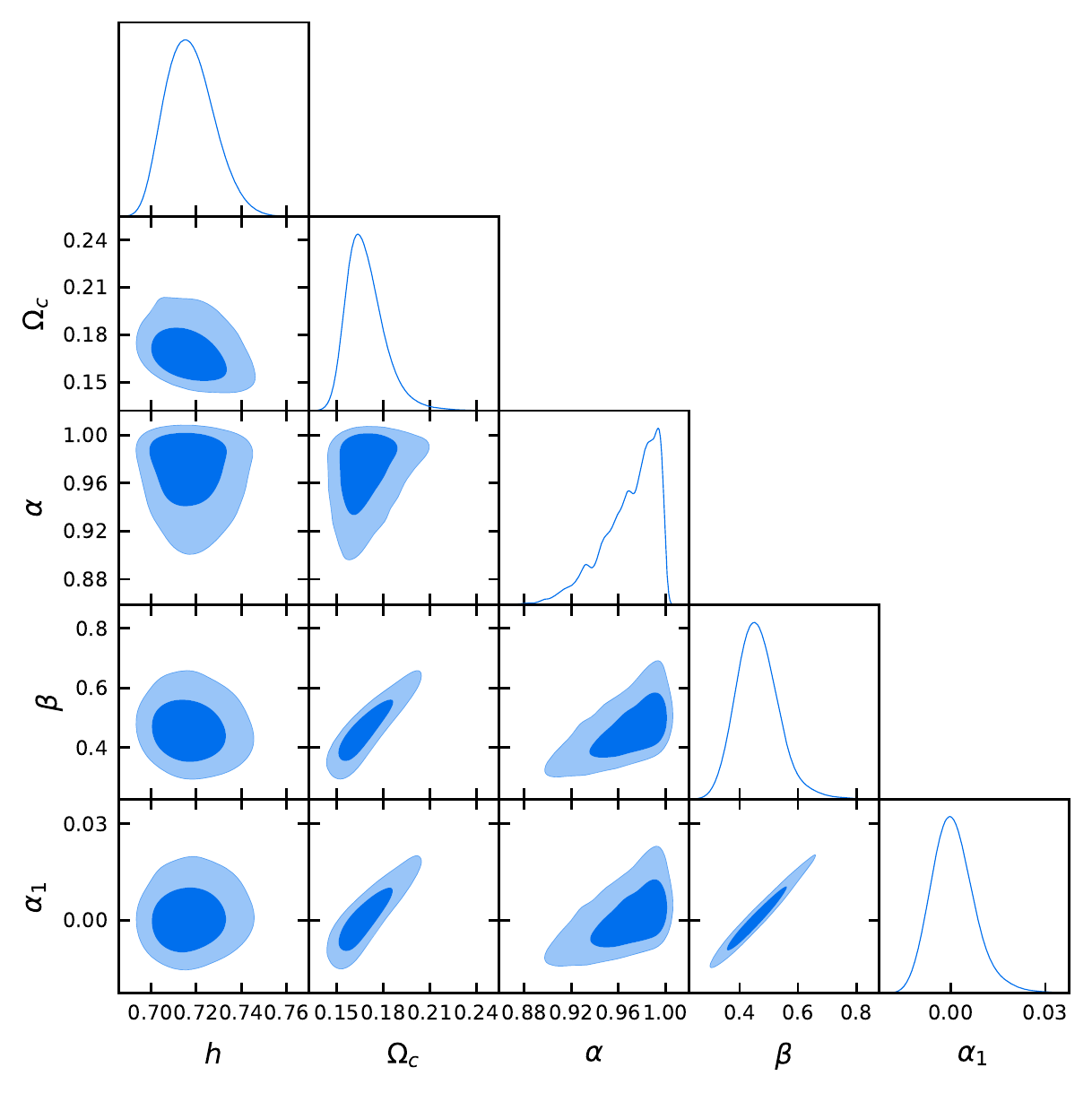}
\caption{Contour plots for interaction model 3 with the $1\sigma$ and $2\sigma$ regions. We considered the full joint analysis with Pantheon + BAO + CC + CMB.} \label{Model3}
\end{figure}

\begin{figure*}[ht]
\centering
 \includegraphics[width=0.8\textwidth]{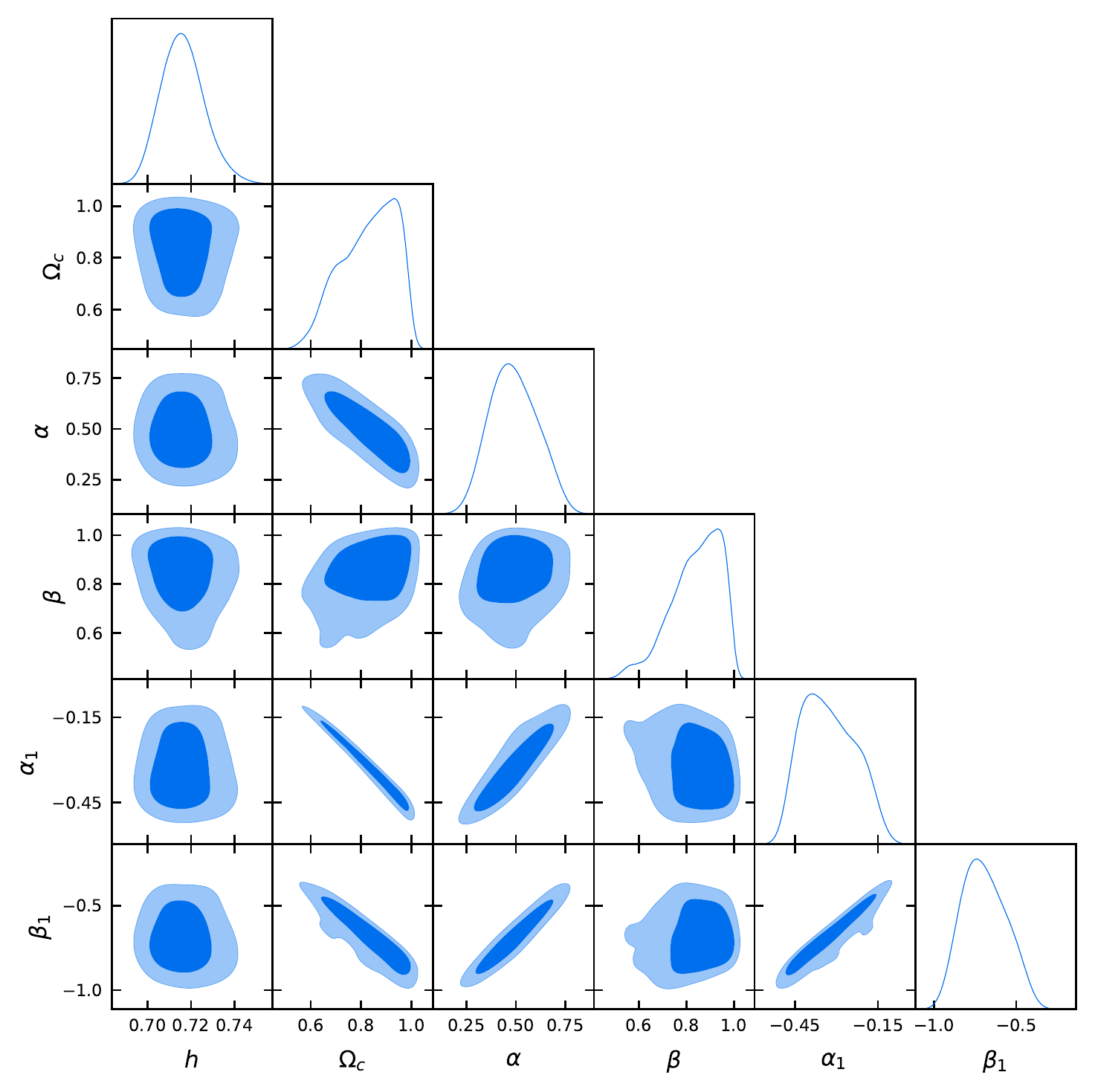}
\caption{Contour plots for interaction model 4 with the $1\sigma$ and $2\sigma$ regions. We considered the full joint analysis with Pantheon + BAO + CC + CMB.} \label{Model4}
\end{figure*}

In figures~\ref{Model1},~\ref{Model2},~\ref{Model3}, and~\ref{Model4} we show the contours of 68.3$\%$ and 95.4$\%$ confidence levels  in our analysis, corresponding to IMHRDE 1, 2, 3, 4, respectively.

Finally, from Eq. \eqref{4FluidosSolrinfinito} we can evaluate the performance of the IMHRDE models in alleviating the coincidence problem. By considering the best fit values for the parameters (table \ref{tab3}) we notice that IHRDEM 1 and 4 alleviate the coincidence problem (the coincidence parameter tends to a positive constant in the future). For model IHRDE 3 we notice that $r_{\infty}$ tends to a negative constant, this is because for a given redshift the dark energy density becomes negative in this scenario. For IHRDE 2 the coincidence problem is not alleviated. 

\begin{landscape}
\begin{center}
\begin{table}
\centering
{\scriptsize{
\caption{Best fit parameters for the analysis with the Pantheon sample. The last two columns show the logarithm of
the Bayes factor ($\ln B$) and the interpretation of the strength of the evidence. Note that $\ln B < -1$ favors the $\Lambda$CDM model.}\label{tab2}
\begin{tabular}{|c|c|c|c|c|c|c|c|c|c|c|}
\hline
 &  &  &  &  &  &  &  & & \\
Model & h & $\Omega_c$ & $\alpha$ & $\beta$ & $\alpha_1$ & $\beta_1$ & $\ln E$ & $\ln B$ & Interpretation \\
 &  &  &  &  &  &  &  & & \\
\hline
 &  &  &  &  &  &  &  & & \\
$\Lambda$CDM & $0.735\pm 0.015$ & $0.257\pm 0.022$ & - & - & - & - & $-521.610\pm0.013$ & - & - \\
 &  &  &  &  &  &  &  & & \\
HRDE&  $0.735\pm 0.015$ & $0.182^{+0.072}_{-0.059}$ & - & $0.502^{+0.065}_{-0.082}$& - & - & $-523.578\pm0.019$ & $-1.968\pm0.023$ & Weak \\
 &  &  &  &  &  &  &  & & \\
MHRDE&  $0.735\pm 0.014$ & $0.46^{+0.14}_{-0.20}$ & $0.754^{+0.24}_{-0.079}$ & $0.749^{+0.24}_{-0.090}$& - & - & $-522.490\pm0.293$ & $-0.880\pm0.293$ & Inconclusive \\
 &  &  &  &  &  &  &  & & \\
IMHRDE 1 & $0.736\pm 0.014$ & $0.58^{+0.14}_{-0.26}$ & $0.63^{+0.31}_{-0.15}$ & $0.813^{+0.18}_{-0.064}$ & - & $0.17^{+0.71}_{-0.37}$ & $-523.101\pm0.054$ & $-1.491 \pm 0.056$ & Weak \\
 &  &  &  &  &  &  &  & & \\
IMHRDE 2 & $0.735\pm 0.014$ & $0.58^{+0.23}_{-0.26}$ & $0.64^{+0.33}_{-0.16}$ & $0.70^{+0.28}_{-0.12}$ & $-0.17^{+0.17}_{-0.21}$ & - & $-523.635\pm0.041$ & $-2.025 \pm 0.043$ & Weak \\
 &  &  &  &  &  &  &  & & \\
IMHRDE 3 & $ 0.735\pm 0.014 $ & $ 0.65^{+0.29}_{-0.17} $ & $ 0.58^{+0.29}_{-0.24} $ & $ 0.72^{+0.26}_{-0.10} $ & $ -0.17^{+0.16}_{-0.19} $ & $ -0.17^{+0.16}_{-0.19} $ & $-523.883\pm0.085$ & $-2.273\pm 0.086$ & Weak \\
 &  &  &  &  &  &  &  & & \\
IMHRDE 4 & $0.736\pm 0.014$ & $ 0.60^{+0.26}_{-0.23}$ & $0.62^{+0.31}_{-0.19}$ & $0.72^{+0.25}_{-0.11}$ & $-0.20\pm{0.26}$ & $0.16^{+0.71}_{-0.41}$ & $-523.765\pm0.068$ 
& $-2.155\pm 0.069$ & Weak \\
 &  &  &  &  &  &  &  & & \\
\hline
\end{tabular}
}}
\end{table}
\end{center}
\end{landscape}

\begin{landscape}
\begin{center}
\begin{table}
{\scriptsize{
\caption{Best fit parameters for the joint analysis with the dataset Pantheon + BAO + CC + CMB. The last two columns show the logarithm of the Bayes factor ($\ln B$) and the interpretation of the strength of the evidence. Note that $\ln B < -1$ favors the $\Lambda$CDM model.}\label{tab3}
\begin{tabular}{|c|c|c|c|c|c|c|c|c|c|c|}
\hline
 &  &  &  &  &  &  &  & & \\
Model & $h$ & $\Omega_c$ & $\alpha$ & $\beta$ & $\alpha_1$ & $\beta_1$ & $\ln E$ & $\ln B$ & Interpretation \\
 &  &  &  &  &  &  &  & & \\
\hline
 &  &  &  &  &  &  &  & & \\
$\Lambda$CDM & $0.6907\pm0.0050$ & $0.2453\pm0.0063$ & - & - & - & - & $-541.229\pm0.010$ & - & - \\
 &  &  &  &  &  &  &  & & \\
HRDE & $0.6951^{+0.0063}_{-0.0076}$ & $0.1743\pm 0.0053$ & - & $0.488\pm 0.014$ & - & - & $-544.438   \pm0.074$ & $-3.209\pm0.075$ & Moderate \\
 &  &  &  &  &  &  &  & & \\
MHRDE & $0.716^{+0.010}_{-0.012}$ &  $0.1660\pm 0.0061$& $0.974^{+0.024}_{-0.0088}$ & $0.454\pm 0.016$ & - & - & $-545.526   \pm0.040$ & $-4.297\pm0.041$ & Moderate \\
 &  &  &  &  &  &  &  & & \\
IMHRDE 1 & $0.7155^{+0.0088}_{-0.012}$ & $0.365^{+0.016}_{-0.030}$ & $0.895^{+0.10}_{-0.039} $ & $0.668^{+0.092}_{-0.12}$ & - & $-0.177^{+0.090}_{-0.042}$ & $-547.346\pm0.021$ & $-6.117\pm0.023$ & Strong \\ 
 &  &  &  &  &  &  &  & & \\
IMHRDE 2 & $0.7182^{+0.0083}_{-0.011}$ & $0.1734^{+0.0092}_{-0 .0241}$ & $0.972^{+0.027}_{-0.0090} $ & $0.476^{+0.058}_{-0.096}$ & $0.001^{+0.012}_{-0.0099}$ & - & $-549.714\pm0.043$ & $-8.485\pm0.044$ & Strong \\ 
 &  &  &  &  &  &  &  & & \\
IMHRDE 3 & $0.7173^{+0.0091}_{-0.012}$ & $ 0.1689^{+0.0082}_{-0.014} $ & $ 0.970^{+0.030}_{-0.0099}$ & $ 0.462^{+0.062}_{-0.077}$ & $ 0.0009^{+0.0057}_{-0.0075} $ & $ 0.0009^{+0.0057}_{-0.0075} $ & $-550.249\pm0.066$ & $-8.796\pm0.067$ & Strong \\
 &  &  &  &  &  &  &  & & \\
IMHRDE 4 & $0.7162^{+0.0088}_{-0.011} $ & $0.831^{+0.15}_{-0.072}    $ & $0.49^{+0.11}_{-0.14}      $ & $0.842^{+0.14}_{-0.063}    $ & $-0.328^{+0.082}_{-0.12}   $ & $-0.69^{+0.12}_{-0.16}     $ & $-549.161\pm0.043$ & $-7.932\pm 0.044$ & Strong \\ 
 &  &  &  &  &  &  &  & & \\
\hline
\end{tabular}
}}
\end{table}
\end{center}
\end{landscape}

\section{Final Remarks}\label{remarks}
In this work we have studied interacting modified holographic Ricci dark energy models, where linear interactions are considered. We have found analytical solutions to these scenarios (see Eq. \eqref{Hz}) and we have performed a bayesian model selection analysis. The bayesian comparison is performed with the combination of background data SNe + CC + BAO + CMB (see section \ref{data}) and the fiducial model is assumed to be $\Lambda$CDM. Our results indicate that there is strong evidence against the IMHRDE models studied, this conclusion is consistent with several studies where holographic dark energy models have been contrasted with background data \cite{Li:2009bn,delCampo:2011jp,Wang:2010kwa,Fu:2011ab,Feng:2016djj,George:2019vko,Akhlaghi:2018knk}.\\

{\it Acknowledgements.} AC was partially support by Direcci\'on de Investigaci\'on Universidad del B\'io-B\'io through grant no. GI-172309/C. CRB wants to thank the financial support of Direcci\'on de Postgrado and Direcci\'on de Investigaci\'on Universidad del B\'io-B\'io.


\begin{thebibliography}{}
\bibitem{Weinberg:2012es}
D.~H.~Weinberg, M.~J.~Mortonson, D.~J.~Eisenstein, C.~Hirata, A.~G.~Riess and E.~Rozo,
%``Observational Probes of Cosmic Acceleration,''
Phys. Rept. \textbf{530}, 87-255 (2013)
%doi:10.1016/j.physrep.2013.05.001
%[arXiv:1201.2434 [astro-ph.CO]].

\bibitem{Weinberg:1988cp}
S.~Weinberg,
%``The Cosmological Constant Problem,''
Rev. Mod. Phys. \textbf{61}, 1-23 (1989)
%doi:10.1103/RevModPhys.61.1

\bibitem{Weinberg:2000yb}
S.~Weinberg,
%``The Cosmological constant problems,''
[arXiv:astro-ph/0005265 [astro-ph]].

\bibitem{Chimento:2003iea}
L.~P.~Chimento, A.~S.~Jakubi, D.~Pavon and W.~Zimdahl,
%``Interacting quintessence solution to the coincidence problem,''
Phys. Rev. D \textbf{67}, 083513 (2003)
%doi:10.1103/PhysRevD.67.083513
%[arXiv:astro-ph/0303145 [astro-ph]].

\bibitem{Guo:2004vg}
Z.~K.~Guo and Y.~Z.~Zhang,
%``Interacting phantom energy,''
Phys. Rev. D \textbf{71}, 023501 (2005)
%doi:10.1103/PhysRevD.71.023501
%[arXiv:astro-ph/0411524 [astro-ph]].

\bibitem{Pavon:2005yx}
D.~Pavon and W.~Zimdahl,
%``Holographic dark energy and cosmic coincidence,''
Phys. Lett. B \textbf{628}, 206-210 (2005)
doi:10.1016/j.physletb.2005.08.134
[arXiv:gr-qc/0505020 [gr-qc]].

\bibitem{Verde:2019ivm}
L.~Verde, T.~Treu and A.~Riess,
``Tensions between the Early and the Late Universe,''
%doi:10.1038/s41550-019-0902-0
[arXiv:1907.10625 [astro-ph.CO]].

\bibitem{Riess:2016jrr} 
  A.~G.~Riess {\it et al.},
  %``A 2.4% Determination of the Local Value of the Hubble Constant,''
  Astrophys.\ J.\  {\bf 826}, no. 1, 56 (2016)
  %doi:10.3847/0004-637X/826/1/56
 % [arXiv:1604.01424 [astro-ph.CO]];

\bibitem{Riess:2011yx} 
  A.~G.~Riess {\it et al.},
  %``A 3% Solution: Determination of the Hubble Constant with the Hubble Space Telescope and Wide Field Camera 3,''
  Astrophys.\ J.\  {\bf 730}, 119 (2011)
  Erratum: [Astrophys.\ J.\  {\bf 732}, 129 (2011)]
  %doi:10.1088/0004-637X/732/2/129, 10.1088/0004-637X/730/2/119
 % [arXiv:1103.2976 [astro-ph.CO]].

\bibitem{Copeland:2006wr}
E.~J.~Copeland, M.~Sami and S.~Tsujikawa,
%``Dynamics of dark energy,''
Int. J. Mod. Phys. D \textbf{15}, 1753-1936 (2006)
%doi:10.1142/S021827180600942X
%[arXiv:hep-th/0603057 [hep-th]].

\bibitem{Yoo:2012ug}
J.~Yoo and Y.~Watanabe,
%``Theoretical Models of Dark Energy,''
Int. J. Mod. Phys. D \textbf{21}, 1230002 (2012)
%doi:10.1142/S0218271812300029
%[arXiv:1212.4726 [astro-ph.CO]].

\bibitem{Bahamonde:2017ize}
S.~Bahamonde, C.~G.~Böhmer, S.~Carloni, E.~J.~Copeland, W.~Fang and N.~Tamanini,
%``Dynamical systems applied to cosmology: dark energy and modified gravity,''
Phys. Rept. \textbf{775-777}, 1-122 (2018)
%doi:10.1016/j.physrep.2018.09.001
%[arXiv:1712.03107 [gr-qc]].

\bibitem{Wang:2016och}
S.~Wang, Y.~Wang and M.~Li,
%``Holographic Dark Energy,''
Phys. Rept. \textbf{696}, 1-57 (2017)
%doi:10.1016/j.physrep.2017.06.003
%[arXiv:1612.00345 [astro-ph.CO]].

\bibitem{Cohen:1998zx}
A.~G.~Cohen, D.~B.~Kaplan and A.~E.~Nelson,
%``Effective field theory, black holes, and the cosmological constant,''
Phys. Rev. Lett. \textbf{82}, 4971-4974 (1999)
%doi:10.1103/PhysRevLett.82.4971
%[arXiv:hep-th/9803132 [hep-th]].

\bibitem{Li:2004rb}
M.~Li,
%``A Model of holographic dark energy,''
Phys. Lett. B \textbf{603}, 1 (2004)
%doi:10.1016/j.physletb.2004.10.014
%[arXiv:hep-th/0403127 [hep-th]].

\bibitem{Li:2010cj}
M.~Li and Y.~Wang,
%``Quantum UV/IR Relations and Holographic Dark Energy from Entropic Force,''
Phys. Lett. B \textbf{687}, 243-247 (2010)
%doi:10.1016/j.physletb.2010.03.042
%[arXiv:1001.4466 [hep-th]].

\bibitem{Hsu:2004ri}
S.~D.~Hsu,
%``Entropy bounds and dark energy,''
Phys. Lett. B \textbf{594}, 13-16 (2004)
%doi:10.1016/j.physletb.2004.05.020
%[arXiv:hep-th/0403052 [hep-th]].

\bibitem{Gao:2007ep}
C.~Gao, F.~Wu, X.~Chen and Y.~G.~Shen,
%``A Holographic Dark Energy Model from Ricci Scalar Curvature,''
Phys. Rev. D \textbf{79}, 043511 (2009)
%doi:10.1103/PhysRevD.79.043511
%[arXiv:0712.1394 [astro-ph]].

\bibitem{Cai:2008nk}
R.~G.~Cai, B.~Hu and Y.~Zhang,
%``Holography, UV/IR Relation, Causal Entropy Bound and Dark Energy,''
Commun. Theor. Phys. \textbf{51}, 954-960 (2009)
doi:10.1088/0253-6102/51/5/39
[arXiv:0812.4504 [hep-th]].

\bibitem{Xu:2008rp}
L.~Xu, W.~Li and J.~Lu,
%``Cosmic Constraint on Ricci Dark Energy Model,''
Mod. Phys. Lett. A \textbf{24}, 1355-1360 (2009)
%doi:10.1142/S0217732309028850
%[arXiv:0810.4730 [astro-ph]].

\bibitem{Feng:2008kz}
C.~J.~Feng,
%``Reconstructing Quintom from Ricci Dark Energy,''
Phys. Lett. B \textbf{672}, 94-97 (2009)
%doi:10.1016/j.physletb.2009.01.022
%[arXiv:0810.2594 [hep-th]].

\bibitem{Feng:2008rs}
C.~J.~Feng,
%``Statefinder Diagnosis for Ricci Dark Energy,''
Phys. Lett. B \textbf{670}, 231-234 (2008)
%doi:10.1016/j.physletb.2008.11.005
%[arXiv:0809.2502 [hep-th]].

\bibitem{Li:2009bn}
M.~Li, X.~D.~Li, S.~Wang and X.~Zhang,
%``Holographic dark energy models: A comparison from the latest observational data,''
JCAP \textbf{06}, 036 (2009)
%doi:10.1088/1475-7516/2009/06/036
%[arXiv:0904.0928 [astro-ph.CO]].

%\cite{Zhang:2009un}
\bibitem{Zhang:2009un} 
  X.~Zhang,
  %``Holographic Ricci dark energy: Current observational constraints, quintom feature, and the reconstruction of scalar-field dark energy,''
  Phys.\ Rev.\ D {\bf 79}, 103509 (2009)
% doi:10.1103/PhysRevD.79.103509
% [arXiv:0901.2262 [astro-ph.CO]].
  
\bibitem{Lepe:2010vh}
S.~Lepe and F.~Pena,
%``Crossing the phantom divide with Ricci-like holographic dark energy,''
Eur. Phys. J. C \textbf{69}, 575-579 (2010)
%doi:10.1140/epjc/s10052-010-1427-y
%[arXiv:1005.2180 [hep-th]].

\bibitem{Kim:2008ej}
K.~Y.~Kim, H.~W.~Lee and Y.~S.~Myung,
%``On the Ricci dark energy model,''
Gen. Rel. Grav. \textbf{43}, 1095-1101 (2011)
%doi:10.1007/s10714-010-0941-4
%[arXiv:0812.4098 [gr-qc]].

\bibitem{delCampo:2011jp}
S.~del Campo, J.~Fabris, R.~Herrera and W.~Zimdahl,
%``On holographic dark-energy models,''
Phys. Rev. D \textbf{83}, 123006 (2011)
%doi:10.1103/PhysRevD.83.123006
%[arXiv:1103.3441 [astro-ph.CO]].

\bibitem{delCampo:2013hka}
S.~del Campo, J.~C.~Fabris, R.~Herrera and W.~Zimdahl,
%``Cosmology with Ricci dark energy,''
Phys. Rev. D \textbf{87}, no.12, 123002 (2013)
%doi:10.1103/PhysRevD.87.123002
%[arXiv:1303.3436 [astro-ph.CO]].

\bibitem{Granda:2008dk}
L.~Granda and A.~Oliveros,
%``Infrared cut-off proposal for the Holographic density,''
Phys. Lett. B \textbf{669}, 275-277 (2008)
%doi:10.1016/j.physletb.2008.10.017
%[arXiv:0810.3149 [gr-qc]].

\bibitem{Granda:2008tm}
L.~Granda and A.~Oliveros,
%``New infrared cut-off for the holographic scalar fields models of dark energy,''
Phys. Lett. B \textbf{671}, 199-202 (2009)
%doi:10.1016/j.physletb.2008.12.025
%[arXiv:0810.3663 [gr-qc]].

\bibitem{Granda:2009xu}
L.~Granda and A.~Oliveros,
%``Holographic reconstruction of the k-essence and dilaton models,''
[arXiv:0901.0561 [hep-th]].

\bibitem{Granda:2009di}
L.~Granda, W.~Cardona and A.~Oliveros,
%``Current observational constraints on holographic dark energy model,''
[arXiv:0910.0778 [hep-th]].

\bibitem{Wang:2010kwa}
Y.~Wang and L.~Xu,
%``Current Observational Constraints to Holographic Dark Energy Model with New Infrared cut-off via Markov Chain Monte Carlo Method,''
Phys. Rev. D \textbf{81}, 083523 (2010)
%doi:10.1103/PhysRevD.81.083523
%[arXiv:1004.3340 [astro-ph.CO]]. 

\bibitem{Mathew:2012md}
T.~K.~Mathew, J.~Suresh and D.~Divakaran,
%``Modified holographic Ricci dark energy model and statefinder diagnosis in flat universe,''
Int. J. Mod. Phys. D \textbf{22}, 1350056 (2013)
%doi:10.1142/S0218271813500569
%[arXiv:1207.5886 [astro-ph.CO]].

\bibitem{Hu:2006ar}
B.~Hu and Y.~Ling,
%``Interacting dark energy, holographic principle and coincidence problem,''
Phys. Rev. D \textbf{73}, 123510 (2006)
%doi:10.1103/PhysRevD.73.123510
%[arXiv:hep-th/0601093 [hep-th]].

\bibitem{Fu:2011ab}
T.~F.~Fu, J.~F.~Zhang, J.~Q.~Chen and X.~Zhang,
%``Holographic Ricci dark energy: Interacting model and cosmological constraints,''
Eur. Phys. J. C \textbf{72}, 1932 (2012)
%doi:10.1140/epjc/s10052-012-1932-2
%[arXiv:1112.2350 [astro-ph.CO]].

\bibitem{Chimento:2011dw}
L.~P.~Chimento, M.~I.~Forte and M.~G.~Richarte,
%``Self-interacting holographic dark energy,''
Mod. Phys. Lett. A \textbf{28}, 1250235 (2013)
%doi:10.1142/S0217732312502355
%[arXiv:1106.0781 [astro-ph.CO]].

\bibitem{Chimento:2011pk}
L.~P.~Chimento and M.~G.~Richarte,
%``Interacting dark matter and modified holographic Ricci dark energy induce a relaxed Chaplygin gas,''
Phys. Rev. D \textbf{84}, 123507 (2011)
%doi:10.1103/PhysRevD.84.123507
%[arXiv:1107.4816 [astro-ph.CO]].

\bibitem{Chimento:2012zz}
L.~P.~Chimento and M.~G.~Richarte,
%``Interacting dark matter and modified holographic Ricci dark energy plus a noninteracting cosmic component,''
Phys. Rev. D \textbf{85}, 127301 (2012)
%doi:10.1103/PhysRevD.85.127301
%[arXiv:1207.1492 [astro-ph.CO]].

\bibitem{Chimento:2012hn} %chimentointeracciongeneral 
  L.~P.~Chimento, M.~I.~Forte and M.~G.~Richarte,
  %``Holographic dark energy linearly interacting with dark matter,''
  AIP Conf.\ Proc.\  {\bf 1471}, 39 (2012)
  %doi:10.1063/1.4756809
  %[arXiv:1206.0179 [gr-qc]].

\bibitem{Chimento:2013se}
L.~P.~Chimento, M.~Forte and M.~G.~Richarte,
%``Modified holographic Ricci dark energy coupled to interacting dark matter and a non interacting baryonic component,''
Eur. Phys. J. C \textbf{73}, no.1, 2285 (2013)
%doi:10.1140/epjc/s10052-013-2285-1
%[arXiv:1301.2737 [gr-qc]].

\bibitem{P:2013cmq}
P.~Pankunni and T.~K.~Mathew,
%``Interacting modified holographic Ricci dark energy model and statefinder diagnosis in flat universe,''
Int. J. Mod. Phys. D \textbf{23}, 1450024 (2014)
%doi:10.1142/S0218271814500242
%[arXiv:1309.3136 [astro-ph.CO]].

\bibitem{Arevalo:2013tta}
F.~Arevalo, P.~Cifuentes, S.~Lepe and F.~Peña,
%``Interacting Ricci-like holographic dark energy,''
Astrophys. Space Sci. \textbf{352}, 899-907 (2014)
%doi:10.1007/s10509-014-1946-3
%[arXiv:1308.5007 [gr-qc]].

\bibitem{Chimento:2013qja}
L.~P.~Chimento and M.~G.~Richarte,
%``Dark radiation and dark matter coupled to holographic Ricci dark energy,''
Eur. Phys. J. C \textbf{73}, no.4, 2352 (2013)
%doi:10.1140/epjc/s10052-013-2352-7
%[arXiv:1303.3356 [gr-qc]].

\bibitem{Oliveros:2014kla}
A.~Oliveros and M.~A.~Acero,
%``New holographic dark energy model with non-linear interaction,''
Astrophys. Space Sci. \textbf{357}, no.1, 12 (2015)
%doi:10.1007/s10509-015-2310-y
%[arXiv:1412.7244 [hep-th]].

\bibitem{Som:2014hja}
S.~Som and A.~Sil,
%``Interacting holographic dark energy models: a general approach,''
Astrophys. Space Sci. \textbf{352}, 867-875 (2014)
%doi:10.1007/s10509-014-1926-7
%[arXiv:1412.0526 [gr-qc]].

\bibitem{Mahata:2015nga}
N.~Mahata and S.~Chakraborty,
%``A dynamical system analysis of holographic dark energy models with different IR cutoff,''
Mod. Phys. Lett. A \textbf{30}, no.27, 1550134 (2015)
%doi:10.1142/S0217732315501345
%[arXiv:1511.07955 [gr-qc]].

\bibitem{Pan:2014afa}
S.~Pan and S.~Chakraborty,
%``A cosmographic analysis of holographic dark energy models,''
Int. J. Mod. Phys. D \textbf{23}, no.11, 1450092 (2014)
%doi:10.1142/S0218271814500928
%[arXiv:1410.8281 [gr-qc]].

\bibitem{Arevalo:2014zoa}
F.~Arevalo, P.~Cifuentes and F.~Pena,
%``Thermodynamics of interacting holographic dark energy,''
Astrophys. Space Sci. \textbf{361}, no.1, 45 (2016)
%doi:10.1007/s10509-015-2634-7
%[arXiv:1408.5118 [hep-th]].

\bibitem{Lepe:2015qhq}
S.~Lepe and F.~Peña,
%``Interacting cosmic fluids and phase transitions under a holographic modeling for dark energy,''
Eur. Phys. J. C \textbf{76}, no.9, 507 (2016)
%doi:10.1140/epjc/s10052-016-4347-7
%[arXiv:1511.07186 [gr-qc]].

\bibitem{Zadeh:2016vgc}
M.~A.~Zadeh, A.~Sheykhi and H.~Moradpour,
%``Holographic dark energy with the sign-changeable interaction term,''
Int. J. Mod. Phys. D \textbf{26}, no.08, 1750080 (2017)
%doi:10.1142/S0218271817500808
%[arXiv:1610.08093 [gr-qc]].

\bibitem{Herrera:2016uci}
R.~Herrera, W.~Hipolito-Ricaldi and N.~Videla,
%``Instability in interacting dark sector: An appropriate Holographic Ricci dark energy model,''
JCAP \textbf{08}, 065 (2016)
%doi:10.1088/1475-7516/2016/08/065
%[arXiv:1607.01806 [gr-qc]].
 
\bibitem{Feng:2016djj}
L.~Feng and X.~Zhang,
%``Revisit of the interacting holographic dark energy model after Planck 2015,''
JCAP \textbf{08}, 072 (2016)
%doi:10.1088/1475-7516/2016/08/072
%[arXiv:1607.05567 [astro-ph.CO]].

\bibitem{George:2019vko}
P.~George and T.~K.~Mathew,
%``Bayesian analysis of running holographic Ricci dark energy,''
[arXiv:1906.08532 [gr-qc]].

\bibitem{Akhlaghi:2018knk}
I.~Akhlaghi, M.~Malekjani, S.~Basilakos and H.~Haghi,
%``Model selection and constraints from Holographic dark energy scenarios,''
Mon. Not. Roy. Astron. Soc. \textbf{477}, no.3, 3659-3671 (2018)
%doi:10.1093/mnras/sty903
%[arXiv:1804.02989 [gr-qc]].

\bibitem{Arevalo:2016epc}
F.~Arevalo, A.~Cid and J.~Moya,
%``AIC and BIC for cosmological interacting scenarios,''
Eur. Phys. J. C \textbf{77}, no.8, 565 (2017)
%doi:10.1140/epjc/s10052-017-5128-7
%[arXiv:1610.09330 [astro-ph.CO]].

\bibitem{Santos:2016sti}
B.~Santos, N.~C.~Devi and J.~Alcaniz,
%``Bayesian comparison of nonstandard cosmologies using type Ia supernovae and BAO data,''
Phys. Rev. D \textbf{95}, no.12, 123514 (2017)
%doi:10.1103/PhysRevD.95.123514
%[arXiv:1603.06563 [astro-ph.CO]].

\bibitem{Heavens:2017hkr}
A.~Heavens, Y.~Fantaye, E.~Sellentin, H.~Eggers, Z.~Hosenie, S.~Kroon and A.~Mootoovaloo,
%``No evidence for extensions to the standard cosmological model,''
Phys. Rev. Lett. \textbf{119}, no.10, 101301 (2017)
%doi:10.1103/PhysRevLett.119.101301
%[arXiv:1704.03467 [astro-ph.CO]].

\bibitem{SantosdaCosta:2017ctv}
S.~Santos da Costa, M.~Benetti and J.~Alcaniz,
%``A Bayesian analysis of inflationary primordial spectrum models using Planck data,''
JCAP \textbf{03}, 004 (2018)
%doi:10.1088/1475-7516/2018/03/004
%[arXiv:1710.01613 [astro-ph.CO]].

\bibitem{Andrade:2017iam}
U.~Andrade, C.~Bengaly, J.~Alcaniz and B.~Santos,
%``Isotropy of low redshift type Ia Supernovae: A Bayesian analysis,''
Phys. Rev. D \textbf{97}, no.8, 083518 (2018)
%doi:10.1103/PhysRevD.97.083518
%[arXiv:1711.10536 [astro-ph.CO]].

\bibitem{Ferreira:2017yby}
T.~Ferreira, C.~Pigozzo, S.~Carneiro and J.~Alcaniz,
%``Interaction in the dark sector: a Bayesian analysis with latest observations,''
[arXiv:1712.05428 [astro-ph.CO]].

\bibitem{Cid:2018ugy}
A.~Cid, B.~Santos, C.~Pigozzo, T.~Ferreira and J.~Alcaniz,
%``Bayesian Comparison of Interacting Scenarios,''
JCAP \textbf{03}, 030 (2019)
%doi:10.1088/1475-7516/2019/03/030
%[arXiv:1805.02107 [astro-ph.CO]].

%%%%%%%%%%%%%%%%%%%%%%%%%%%%%
\bibitem{Moresco:2016mzx} %cc
M.~Moresco {\it et al.},
%``A 6% measurement of the Hubble parameter at $z\sim0.45$: direct evidence of the epoch of cosmic re-acceleration,''
JCAP {\bf 1605}, no. 05, 014 (2016)
%doi:10.1088/1475-7516/2016/05/014
%[arXiv:1601.01701 [astro-ph.CO]].
 
\bibitem{Scolnic:2017caz} % PS
D.~M.~Scolnic {\it et al.},
%``The Complete Light-curve Sample of Spectroscopically Confirmed SNe Ia from Pan-STARRS1 and Cosmological Constraints from the Combined Pantheon Sample,''
Astrophys.\ J.\  {\bf 859}, no. 2, 101 (2018)
%doi:10.3847/1538-4357/aab9bb
%[arXiv:1710.00845 [astro-ph.CO]].

\bibitem{Beutler:2011hx} 
F.~Beutler {\it et al.},
%``The 6dF Galaxy Survey: Baryon Acoustic Oscillations and the Local Hubble Constant,''
Mon.\ Not.\ Roy.\ Astron.\ Soc.\  {\bf 416}, 3017 (2011)
%doi:10.1111/j.1365-2966.2011.19250.x
%[arXiv:1106.3366 [astro-ph.CO]].

\bibitem{Ross:2014qpa} 
A.~J.~Ross, L.~Samushia, C.~Howlett, W.~J.~Percival, A.~Burden and M.~Manera,
%``The clustering of the SDSS DR7 main Galaxy sample – I. A 4 per cent distance measure at $z = 0.15$,''
Mon.\ Not.\ Roy.\ Astron.\ Soc.\  {\bf 449}, no. 1, 835 (2015)
%doi:10.1093/mnras/stv154
%[arXiv:1409.3242 [astro-ph.CO]].

%\bibitem{BAO3} A. J. Cuesta et al., Mon. Not. Roy. Astron. Soc. 457, 1770 (2016), arXiv:1509.06371 [astro-ph.CO].

\bibitem{Ata:2017dya} 
M.~Ata {\it et al.},
%``The clustering of the SDSS-IV extended Baryon Oscillation Spectroscopic Survey DR14 quasar sample: first measurement of baryon acoustic oscillations between redshift 0.8 and 2.2,''
Mon.\ Not.\ Roy.\ Astron.\ Soc.\  {\bf 473}, no. 4, 4773 (2018)
%doi:10.1093/mnras/stx2630
%[arXiv:1705.06373 [astro-ph.CO]].
  
\bibitem{Hou:2018yny} 
J.~Hou {\it et al.},
%``The clustering of the SDSS-IV extended Baryon Oscillation Spectroscopic Survey DR14 quasar sample: anisotropic clustering analysis in configuration-space,''
Mon.\ Not.\ Roy.\ Astron.\ Soc.\  {\bf 480}, no. 2, 2521 (2018) 
%doi:10.1093/mnras/sty1984
%[arXiv:1801.02656 [astro-ph.CO]].  
  
\bibitem{Alam:2016hwk} 
S.~Alam {\it et al.} [BOSS Collaboration],
%``The clustering of galaxies in the completed SDSS-III Baryon Oscillation Spectroscopic Survey: cosmological analysis of the DR12 galaxy sample,''
Mon.\ Not.\ Roy.\ Astron.\ Soc.\  {\bf 470}, no. 3, 2617 (2017)
%doi:10.1093/mnras/stx721
%[arXiv:1607.03155 [astro-ph.CO]].

\bibitem{Bourboux:2017cbm} 
H.~du Mas des Bourboux {\it et al.},
%``Baryon acoustic oscillations from the complete SDSS-III Ly$\alpha$-quasar cross-correlation function at $z=2.4$,''
Astron.\ Astrophys.\  {\bf 608}, A130 (2017)
%doi:10.1051/0004-6361/201731731
%[arXiv:1708.02225 [astro-ph.CO]].
  
\bibitem{Ade:2015rim} 
P.~A.~R.~Ade {\it et al.} [Planck Collaboration],
%``Planck 2015 results. XIV. Dark energy and modified gravity,''
Astron.\ Astrophys.\  {\bf 594}, A14 (2016)
%doi:10.1051/0004-6361/201525814
%[arXiv:1502.01590 [astro-ph.CO]].

\bibitem{Jimenez:2001gg} 
R.~Jimenez and A.~Loeb,
%``Constraining cosmological parameters based on relative galaxy ages,''
Astrophys.\ J.\  {\bf 573}, 37 (2002)
%doi:10.1086/340549
%[astro-ph/0106145].

\bibitem{Verde:2014qea}
L.~Verde, P.~Protopapas and R.~Jimenez,
%``The expansion rate of the intermediate Universe in light of Planck,''
Phys.\ Dark Univ.\  {\bf 5-6}, 307 (2014)
%doi:10.1016/j.dark.2014.09.003
%%[arXiv:1403.2181 [astro-ph.CO]].

\bibitem{Eisenstein:2005su} 
D.~J.~Eisenstein {\it et al.} [SDSS Collaboration],
%``Detection of the Baryon Acoustic Peak in the Large-Scale Correlation Function of SDSS Luminous Red Galaxies,''
Astrophys.\ J.\  {\bf 633}, 560 (2005)
%doi:10.1086/466512
%[astro-ph/0501171].

\bibitem{Eisenstein:1997ik} 
D.~J.~Eisenstein and W.~Hu,
%``Baryonic features in the matter transfer function,''
Astrophys.\ J.\  {\bf 496}, 605 (1998)
%doi:10.1086/305424
%[astro-ph/9709112].
  
\bibitem{Evslin:2017qdn} 
J.~Evslin, A.~A.~Sen and Ruchika,
%``Price of shifting the Hubble constant,''
Phys.\ Rev.\ D {\bf 97}, no. 10, 103511 (2018)
%doi:10.1103/PhysRevD.97.103511
%  [arXiv:1711.01051 [astro-ph.CO]].

\bibitem{Aghanim:2018eyx} 
N.~Aghanim {\it et al.} [Planck Collaboration],
%``Planck 2018 results. VI. Cosmological parameters,''
arXiv:1807.06209 [astro-ph.CO].

\bibitem{Trotta:2008qt} 
R.~Trotta,
%``Bayes in the sky: Bayesian inference and model selection in cosmology,''
Contemp.\ Phys.\  {\bf 49}, 71 (2008)
% doi:10.1080/00107510802066753
%[arXiv:0803.4089 [astro-ph]].
  
\bibitem{Feroz:2007kg}
F.~Feroz and M.~Hobson,
%``Multimodal nested sampling: an efficient and robust alternative to MCMC methods for astronomical data analysis,''
Mon. Not. Roy. Astron. Soc. \textbf{384}, 449 (2008)
%doi:10.1111/j.1365-2966.2007.12353.x
%[arXiv:0704.3704 [astro-ph]].

\bibitem{Feroz:2008xx}
F.~Feroz, M.~Hobson and M.~Bridges,
%``MultiNest: an efficient and robust Bayesian inference tool for cosmology and particle physics,''
Mon. Not. Roy. Astron. Soc. \textbf{398}, 1601-1614 (2009)
%doi:10.1111/j.1365-2966.2009.14548.x
%[arXiv:0809.3437 [astro-ph]].

\bibitem{Riess:2018byc} 
A.~G.~Riess {\it et al.},
%``Milky Way Cepheid Standards for Measuring Cosmic Distances and Application to Gaia DR2: Implications for the Hubble Constant,''
Astrophys.\ J.\  {\bf 861}, no. 2, 126 (2018)
% doi:10.3847/1538-4357/aac82e
%[arXiv:1804.10655 [astro-ph.CO]].
 
\bibitem{Komatsu:2010fb} 
E.~Komatsu {\it et al.} [WMAP Collaboration],
%``Seven-Year Wilkinson Microwave Anisotropy Probe (WMAP) Observations: Cosmological Interpretation,''
Astrophys.\ J.\ Suppl.\  {\bf 192}, 18 (2011)
%doi:10.1088/0067-0049/192/2/18
%[arXiv:1001.4538 [astro-ph.CO]].

\bibitem{Mangano:2005cc} 
G.~Mangano, G.~Miele, S.~Pastor, T.~Pinto, O.~Pisanti and P.~D.~Serpico,
%``Relic neutrino decoupling including flavor oscillations,''
Nucl.\ Phys.\ B {\bf 729}, 221 (2005)
%doi:10.1016/j.nuclphysb.2005.09.041
%[hep-ph/0506164].

\end{thebibliography}
\end{document}